\documentclass[letter,twocolumn]{jpsj3}
\usepackage{bm}
\usepackage{txfonts}

\title{Dimensional Reduction and Odd-Frequency Pairing of the
Checkerboard-Lattice Hubbard Model at 1/4-Filling}

\author{Yuki Yanagi$^{1}$\thanks{E-mail address:
yanagi@issp.u-tokyo.ac.jp}, Yasufumi Yamashita$^{2}$\thanks{E-mail
address: yamasita@ge.ce.nihon-u.ac.jp} and Kazuo Ueda$^{1}$}
\inst{\address{$^{1}$ Institute for Solid State Physics, University of Tokyo,
Kashiwa, Chiba 277-8581, Japan \\
$^{2}$ College of Engineering, Nihon University, Koriyama, Fukushima 963-8642, Japan }}

\abst{The ferromagnetism of the checkerboard-lattice Hubbard model at
quarter filling is one of the few exact ferromagnetic ground states
known in the family of Hubbard models.  When the nearest neighbor
hopping, $t_1$, is negligible compared with the second neighbor one, $t_2$,
the system reduces to a collection of Hubbard chains. We find that the
1D character is surprisingly robust as long as $t_1< t_2$.  This phenomenon
of dimensional reduction due to the geometrical frustration leads to
peculiar magnetic orders with 1D character for intermediate $U$ and is
responsible for odd-frequency superconducting states close to the
magnetic boundary.
}

\kword{checkerboard-lattice Hubbard model, flat-band, ferromagnetism,
dimensional reduction, 
odd-frequency superconductivity}

\begin{document}
\maketitle

\begin{figure}[t]
\begin{center}
\includegraphics[width=8.0cm]{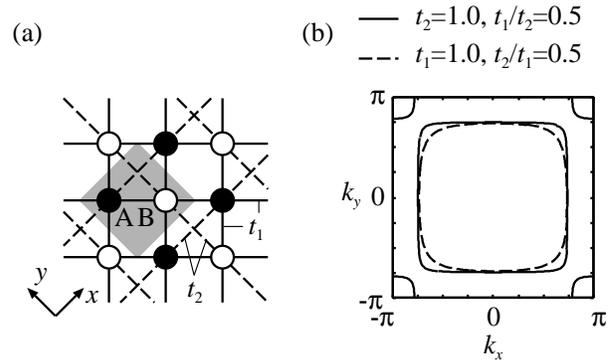}
\caption{(Color online) 
(a) Schematic of the CB-lattice. The solid and dashed lines represent
 $t_1$ and $t_2$ hoppings, respectively. The shaded region is the unit
 cell including two distinct sites A and B denoted by the filled and
 open circles. 
(b) Fermi surfaces for  $t_2=1.0$ and $t_1/t_2=0.5$ (solid lines)
and for $t_1=1.0$ and $t_2/t_1=0.5$ (dashed lines).
\label{fig_checkerboard}}
\end{center}
\end{figure}

Strongly correlated electron systems (SCES) with geometrical frustration
 show a variety of fascinating physics\cite{mielke,kusakabe,fujimoto,yoshioka,tsunetsugu,yamashita,arita,isoda}.
The Hubbard model on the
two-dimensional pyrochlore lattice, called checkerboard-lattice (CB-lattice) Hubbard model, is
 one  typical
example and its intriguing properties, such
as the metal-insulator transition at $1/2$-filling\cite{fujimoto,yoshioka} and the
ferromagnetism at $1/4$-filling\cite{mielke,kusakabe} have been studied
 by many authors. 
The latter is known as the flat-band
ferromagnetism.  Mielke have rigorously proved that the complete
ferromagnetic state is the
ground state of the CB-lattice Hubbard model at $1/4$-filling at the
 symmetric point $t_1=t_2$\cite{mielke}, 
where $t_1$ and $t_2$ are the nearest and next nearest neighbor
hoppings, respectively [see Fig. \ref{fig_checkerboard} (a)]. 
  According to the
numerical calculation, this ferromagnetic
state is extended to the case that $t_1>t_2$\cite{kusakabe}.
The electronic states including the
ferromagnetism for $t_1<t_2$, however, 
have not been investigated so
far. 

In the SCES, the electrons move trying 
to avoid each other in space and
 time due to the strong local Coulomb repulsion 
and it is difficult for them to condense into 
the isotropic pairing state where the on-site and equal-time 
  gap function $\Delta(\bm{r}=\bm{0},t=0)$ is finite with 
  $\bm{r}$ and $t$ being the relative coordinate and time of the Cooper pairs.
 One possible way to stabilize superconductivity in SCES is to
 form  spatially anisotropic Cooper pairs,
 i.e. $\Delta(\bm{r}=\bm{0},t)=0$ and such pairing states are considered
 to be realized in most of the SCES\cite{sigrist}.
An alternative way is
to make the equal-time gap function $\Delta(\bm{r},t=0)$ vanish. In other words, the
gap function has odd-frequency dependence
$\Delta(\bm{r},\omega)=-\Delta(\bm{r},-\omega)$. The odd-frequency pairing
state was first proposed by Berezinskii in the context of the
$^3$He\cite{berezinskii}. 
After his proposal, theoretical studies on the
 odd-frequency pairing have been performed by 
several
 authors\cite{balatsky,coleman,bulut,vojta,fuseya,hotta,shigeta,kusunose}
 and several models have been proposed as play grounds
 for  realization of the
 odd-frequency pairing. 
 Based on these studies, 
 it has been pointed out that 
there are several favorable conditions for
 the odd-frequency pairing: 
(1)  strong retardation effects, 
(2) geometrical frustration and 
(3) one-dimensionality. 
(1) When the retardation effects are strong, 
the effective interaction for the
 odd-frequency pairing is attractive 
for a wide range in frequency
 space and can dominate over the one for the
 even-frequency pairing\cite{kusunose}. 
Strong retardation is realized in a SCES near
 a quantum  critical point (QCP)\cite{fuseya} 
and also in an  electron-phonon system
 with anomalously soft phonons\cite{kusunose}.  
(2)  It has been shown that 
 a geometrical frustration tends to suppress 
 even-frequency  spin-singlet pairing correlation 
and as a result, enhance 
the odd-frequency pairing correlation\cite{vojta,shigeta}.  
(3) In the quasi-one-dimensional system, it has been shown that 
the odd-frequency spin-singlet $p$-wave pairing is
 favored when the one-dimensionality is strong\cite{shigeta}. 
 In this case, the nodes of the gap
 function for the $p$-wave pairing do not intersect the Fermi surfaces
 and the pairing interaction is effective in a wide region of the
 momentum space.  Although it is obvious that the conditions (1) and (2)
 are satisfied in the CB-lattice Hubbard model near the QCP, we will
 show that the  condition (3) is also satisfied in the model.

In this letter, we investigate magnetic properties and 
superconductivity of the CB-lattice Hubbard model at $1/4$-filling
by using the mean field approximation, exact diagonalization (ED) method
  and random phase
approximation (RPA). We show that the charge and magnetic ordered states
with one-dimensional character are observed in the CB-lattice Hubbard
model for $t_1<t_2$. The spin fluctuations also have strong 1D structure
and promote the odd-frequency spin-singlet $p$-wave superconductivity.

This remarkable property of the quasi-one-dimensionality is a
consequence of the geometrical frustration of the present model, which
may be characterized by dimensional reduction. We will show that the
odd-frequency spin-singlet $p$-wave state is stabilized in an extremely
wide parameter range in this model.

\begin{figure}[t]
\begin{center}
\includegraphics[width=8.0cm]{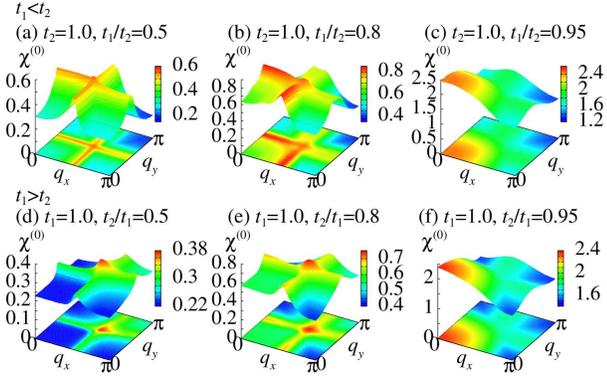}
\caption{(Color online) (a)-(c) $\bm{q}$-dependence of the largest eigenvalue of
 the bare susceptibility $\chi^{(0)}_\mathrm{1st}(\bm{q})$ for $t_1<t_2$
 at $T=0.02t_2$ and
 (d)-(f) that for $t_1>t_2$ at $T=0.02t_1$. \label{fig_chi0}}
\end{center}
\end{figure}

Let us first examine single-electron properties of the CB-lattice Hubbard model at 1/4 filling. When $t_1/t_2=0$, 
there are two 1D bands along $k_x$ and $k_y$ directions, each to be
labeled by A- and B-band. 
Fermi surfaces for the A- and B-bands are respectively 
located along $k_x=\pm 3\pi/4$ and $k_y=\pm 3\pi/4$, 
since both $t_1$ and $t_2$ are positive definite in this paper. 
Looking at the largest eigenvalue of the non-interacting 
paramagnetic susceptibility 
(denoted by $\chi^{(0)}_{\rm 1st}(\bm{q})$ hereafter), 
the perfect nesting vectors of $\bm{q}=(\pi/2,q_y)$ 
and $(q_x,\pi/2)$ give ridge-like structures, 
which are hallmarks of the one dimensionality. 
With increasing $t_1$ from zero toward $t_2$, 
the A- and B-bands start to hybridize and we expect that 
the 1D band structures would be modified significantly. 
Strikingly enough, though, the shape of Fermi surface does not change
much, except for the small regions centered around 
the Fermi-surface crossing points at $(k_x,k_y)=(\pm 3\pi/4,\pm3\pi/4)$ 
[see Fig. \ref{fig_checkerboard}(b)]. 
As a result, a major Fermi surface of a large hole pocket 
around $\Gamma$ point looks square-like and retains 
one dimensionality even for $t_1/t_2=0.8$. 
This is because the 1D-like Fermi surface at 1/4 filling 
lies in the vicinity of the zone boundary of the first Brillouin 
zone where A-B band mixings vanish exactly from the symmetry\cite{tbp}. 
The 1D nature of Fermi surface is also seen 
in $\chi^{(0)}_{\rm 1st}(\bm{q})$ as ridge-like structures 
along $q_x=\pi/2$ and $q_y=\pi/2$ lines as shown in
Figs. \ref{fig_chi0}-(a) and (b). Along the ridge line, 
there is a weak structure and the most divergent mode 
is located at $\bm{q}=(\pi/2,\pi/2)$ for $0<t_1/t_2<0.58$  
and at about $\bm{q}=(\pi/2,0)$ and $(0,\pi/2)$ for $0.66<t_1/t_2<0.91$. 
 In many cases, the geometrical frustration suppresses the $\bm{q}$-dependence
 of the spin fluctuations, leading to suppression of 
 the even-frequency anisotropic pairing\cite{vojta,shigeta}. 
 On the other hand, in the CB-lattice, 
 it produces the above-mentioned robust 
1D features of the single-electron properties and promotes the
odd-frequency pairing as explained later.

On the other hand for $0<t_2/t_1<0.9$, 
the Fermi surface is a single rounded-square electron pocket [Fig. \ref{fig_checkerboard}(b)] and $\chi^{(0)}_{\rm 1st}(\bm{q})$ has an incommensurate peak located at $\bm{q}=(\pi/2+\delta,\pi/2+\delta)$ with $\delta\sim 0.1$. Note that the ridge-like structures around $q_x=\pi/2$ and $q_y=\pi/2$ develop with increasing $t_2/t_1$, see Figs. \ref{fig_chi0}(d) and (e). 
As we approach the Mielke point by changing $t_2/t_1$ or $t_1/t_2$ to
unity, due to the geometrically frustrated hopping processes, the lower
band is flattened and $\chi^{(0)}_{\rm 1st}(\bm{q})$ become
nearly-structureless with a weak peak located at $\bm{q}=(0,0)$ for
$t_1/t_2\sim 1$ as shown in Figs. \ref{fig_chi0} (c) and (f). Near the Mielke point, therefore, strong electron correlation is intrinsic even when $U$ is rather small compared with $t_1$ and $t_2$.

\begin{figure}[t]
\begin{center}
\includegraphics[width=9.0cm]{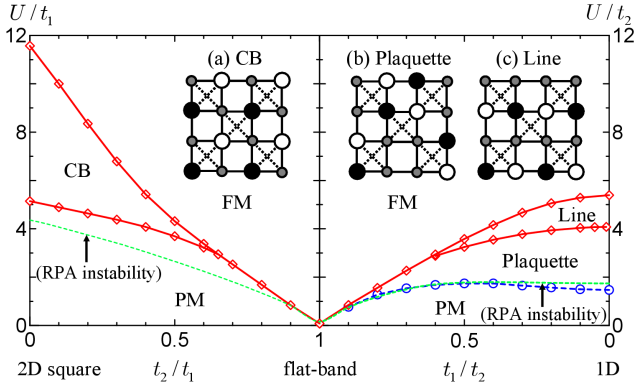}
\caption{(Color online) 
Mean-field phase diagram for $t_2<t_1$ (left-half panel) and $t_2>t_1$ (right). The open circles and diamonds are, respectively, continuous and discontinuous phase-transition points within the present numerical accuracy.
RPA-instability lines are also depicted by the dotted lines. Inset figures (a)-(c) show the unit cells of various charge-ordered AF state, where the circular radius stands for the magnitude of charge density and each color zero (gray), positive (black), or negative (white) spin polarization.
\label{fig_pd0}}
\end{center}
\end{figure}

After switching on the Coulomb interaction $U$, 
within the mean-field approximation, 
the peak modes in $\chi^{(0)}_{\rm 1st}(\bm{q})$ 
become leading magnetic instabilities from the paramagnetic phase 
and second-order phase transition takes place at a certain value 
of $U_c$ for a fixed $t_1/t_2$. When $t_1/t_2=1$, we know that 
the Mielke's ferromagnetism is realized irrespective of $U$, 
meaning that $U_c(t_1/t_2=1)=0$. Exact diagonalization (ED) study 
for 16-site CB-lattice Hubbard model under anti-periodic boundary 
conditions shows that a critical $U_c$ from the paramagnetic ($S=0$) 
to the ferromagnetic ($S=4$) states grows rapidly by changing $t_1/t_2$ 
or $t_2/t_1$ from unity, like $U_c(t_2/t_1=0.8)=4.0537t_1$ 
or $U_c(t_1/t_2=0.8)=4.5105t_2$\cite{kusakabe,tbp}. 
Since the flat-band ferromagnetism would be suppressed 
when the bands become dispersive, we consider 
the magnetic phase diagram in $U$-$t_1$ or $U$-$t_2$ plane 
by applying the mean-field approximation.  
For this purpose, a unit cell is extended to the $2\sqrt{2}\times2\sqrt{2}$ ones 
shown in the inset of Fig. \ref{fig_pd0} and set up charge-modulated
antiferromagnets (AFs) of line-, plaquette-, and checkerboard-type as
well as usual paramagnet (PM) and ferromagnet (FM), for possible
mean-field ground states. The line- and plaquette-AF have
$\bm{q}=(\pi/2,\pi/2)$ spin structure and are made from different
arrangements of classical charge and spin ordered chains along the $t_2$-bond directions at 1/4-filling. 
In the case of $t_2/t_1<1$ and relatively large $U$, the checkerboard-type AF ground state is expected, since the charge modulation in this state are so distributed as to gain kinetic energy through $t_1$ hoppings. These states are consistent with the short-range correlations calculated by ED. 
 Fig. \ref{fig_pd0} shows the magnetic mean-field phase diagram in $U$-$t_1$ (left panel) or $U$-$t_2$ plane (right) as well as the RPA instabilities. The self-consistent energy of the different ordered states are typically calculated for a 32$\times$32 number of enlarged unit cells at $T=0.01t_1$ for $t_2/t_1<1$ or at $T=0.01t_2$ for  $t_1/t_2<1$\cite{tbp}. 
From Fig. \ref{fig_pd0}, one can see that Mielke's ferromagnetic phase
is extended for large-$U$ region and various AF phases emerge between
the paramagnetic and ferromagnetic phases. 
When $0<t_1/t_2<0.2$, the self-consistent energy of the line- and the plaquette-states are almost degenerate owing to the one-dimensionality of the system, which may lead to the discrepancy between the phase boundary estimated by the mean-field calculation at $T=0.01t_2$ and that by RPA at $T=0.02t_2$. It should be also mentioned that, because of the size limitation of the unit cell in the mean-field approximation, we could not reproduce the second-order phase boundary from PM to the incommensurate state found in $\chi^{(0)}_{\rm 1st}$ for $t_1>t_2$. It is natural to expect that there would be successive phase transitions between RPA instability line and the first order transition line shown in Fig. \ref{fig_pd0}. 

Now, let us turn our attention to the superconductivity. 
We investigate the superconductivity by solving the following linearized
Eliashberg equation within the RPA, 
\begin{eqnarray}
\lambda\Delta_{\alpha\beta}(k)&=&-\frac{T}{N}\sum_{k'}
\sum_{\alpha'\beta'}V_{\alpha\beta}(k-k')\ \ \ \ \ \ \nonumber\\
&\times&G^{(0)}_{\alpha'\alpha}(-k')
\Delta_{\alpha'\beta'}(k') G^{(0)}_{\beta'\beta}(k') \label{eq_gap},
\end{eqnarray}
where $G^{(0)}_{\alpha\beta}(k)$ is the noninteracting Green's function, 
$\Delta_{\alpha\beta}(k)$ is the gap function
, $V_{\alpha\beta}(q)$ is the
effective pairing interaction and
$\lambda$ is the eigenvalue of the eigenvalue equation (\ref{eq_gap}) 
which represents the strength of the superconducting
correlation and reaches unity at  $T=T_c$. In the RPA,
$V_{\alpha\beta}(q)$ is given as,
\begin{equation}
\hat{V}(q)=\eta U^2
 \hat{\chi}^s(q)-\frac{1}{2}U^2\hat{\chi}^c(q)+U\label{eq_veff},
\end{equation}
where $\eta=3/2(-1/2)$ for the spin-singlet (-triplet) state and the spin (charge) susceptibility $\hat{\chi}^{s(c)}(q)$ is given as,
$\hat{\chi}^{s(c)}(q)=[\hat{1}-(+)U\hat{\chi}^{(0)}(q)]^{-1}\hat{\chi}^{(0)}(q)$
with the bare susceptibility 
$\chi^{(0)}_{\alpha\beta}(q)=-T/N\sum_k G^{(0)}_{\alpha\beta}(k)G^{(0)}_{\beta\alpha}(k-q)$.
We use the abbreviations $k=(\bm{k},i\varepsilon_n)$ and
$q=(\bm{q},i\omega_m)$, where $\varepsilon_n=(2n+1)\pi T$ and
$\omega_m=2m\pi T$. In the numerical calculations, we use the 
$128\times 128$ $\bm{k}$-meshes in the 1st Brillouin zone and 
$512$ Matsubara frequencies $(-511\pi T \le \varepsilon_n \le 511 \pi T)$ and set
$T=0.02t_1$ and $T=0.02t_2$ for $t_1>t_2$ and $t_1 < t_2$, respectively.

\begin{figure}[t]
\begin{center}
\includegraphics[width=8.0cm]{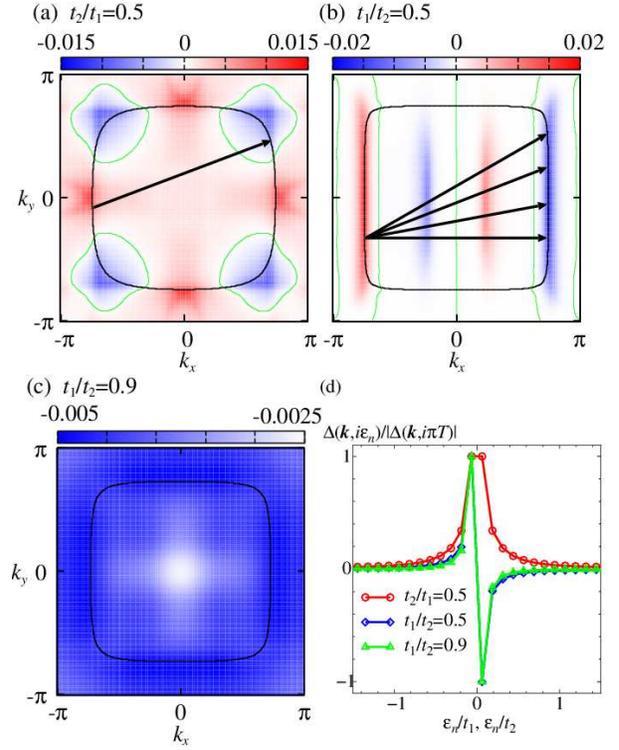}
\caption{(Color online) (a), (b) and (c) $\bm{k}$-dependence of the lower band diagonal
 component of the gap function $\Delta(\bm{k},i\varepsilon_n)$ with the lowest
 Matsubara frequency $\varepsilon_n=\pi T$ for $t_1=1.0$ and $t_2/t_1=0.5$, $t_2=1.0$ and
 $t_1/t_2=0.5$ and $t_2=1.0$ and $t_1/t_2=0.9$, respectively, The
 black and green (gray) lines denote the Fermi surfaces and the nodes of
 the gap function, respectively and the arrows schematically represent the typical
 pair scatterings. (d)
 $\varepsilon_n$-dependence of that on the Fermi
 surface $\bm{k}=$ $\bm{k}_F\sim (3\pi/4,0)$ \label{fig_gap}}
\end{center}
\end{figure}

Fig \ref{fig_gap} shows the $\bm{k}$- and the $\varepsilon_n$-dependence
of the obtained gap functions 
for several typical parameter sets 
in the cases that $t_1<t_2$, $t_1>t_2$ and $t_1\sim t_2$, 
where the values of $U$ are
chosen as the largest eigenvalue $\lambda \sim 1$. 
We note that in the present case, the effective
pairing interaction given in eq. (\ref{eq_veff}) 
can be approximated
by $\hat{V}(q)\sim \eta U^2\hat{\chi}^s (q)$, that is, 
the superconductivity is
driven by the spin fluctuations 
because the spin fluctuations always 
dominate over the charge fluctuations
and are strongly enhanced by the Coulomb interaction $U$ 
when $\lambda\sim 1$. With these in mind, we
discuss the pairing symmetry and mechanism.  
It is found that the pairing symmetry 
for $t_1=1.0$ and $t_2/t_1=0.5$ is even-frequency
spin-singlet extended $s$-wave (ES$s$) as shown in Figs \ref{fig_gap} (a) and (d).
Since the effective pairing interaction 
for the spin-singlet state given in eq. (\ref{eq_veff}) has a sharp peak
at $\bm{q}\sim (\pi/2,\pi/2)$ reflecting the structure of $\hat{\chi}^{(0)}(q)$
shown in Fig. \ref{fig_chi0} (d), strong repulsive pair scatterings with
momentum transfer $\bm{q}\sim (\pi/2,\pi/2)$ take place 
[see Fig. \ref{fig_gap} (a)]. 
Then, the gap function has different signs at the segments on the Fermi surface 
which are connected to each
other by $\bm{q}\sim (\pi/2,\pi/2)$. 

On the other hand, the pairing symmetry 
for $t_2=1.0$ and $t_1/t_2=0.5$ is odd-frequency
spin-singlet $p$-wave (OS$p$) as shown in Figs. \ref{fig_gap} 
(b) and (d). Since the effective pairing interaction 
for the spin-singlet state given in eq. (\ref{eq_veff}) 
has ridge-like structures along  $q_x=\pi/2$ and $q_y=\pi/2$, 
strong repulsive pair scatterings with
momentum transfer $\bm{q}\sim (\pi/2,q_y)$ and 
$\bm{q}\sim (q_x,\pi/2)$ are important. 
Therefore, electrons on the Fermi surface
with $k_x \sim -3\pi/4$ ($k_y \sim -3\pi/4$) 
are scattered to the other section of the Fermi
surface with $k_x \sim 3\pi/4$ ($k_y \sim -3\pi/4$) 
and vice versa  [see Fig. \ref{fig_gap} (b)]. 
Then, the gap function has sign
change between the sections of the Fermi surface which are connected to each
other by $\bm{q}\sim (\pi/2,q_y)$ and the resulting $k$-dependence of
the gap function is $p_x$-wave. We note that the $p_x$- and
$p_y$-wave states are degenerate in the CB-lattice. 
Since the relation $\Delta(\bm{k},i\varepsilon_n)=\Delta(-\bm{k},-i\varepsilon_n)$ has to
be hold for the spin-singlet state,  
the frequency-dependence of the gap function is odd.

In contrast to the above-mentioned two cases, near the Mielke point, 
the spin-triplet and spatially isotropic state, i.e., 
the odd-frequency spin-triplet $s$-wave (OT$s$) pairing is realized as
shown in Figs. \ref{fig_gap} (c) and (d).
Near the Mielke point, the lower bandwidth is narrow and
 the $\bm{q}$-dependence of $\hat{\chi}^{(0)}(q)$ is weak as shown in
 Figs. \ref{fig_chi0} (c) and (f).  
Then, the $\bm{q}$-dependence of the effective pairing interaction 
for the spin-triplet state given in eq. (\ref{eq_veff}) is also weak
 and $V(q)<0$ for any $\bm{q}$ which lead to strong 
attractive pair scatterings. Thus, the gap function shows no sign
change and  is of $s$-wave. 
Since $\Delta(\bm{k},i\varepsilon_n)=-\Delta(-\bm{k},-i\varepsilon_n)$ has to
be hold for the spin-triplet state,  
the frequency-dependence of the gap function is odd. Two distinct
odd-frequency pairing states, the OS$p$ and OT$s$ states are realized
due to the spin fluctuations with quite different features characterized
by the strong one-dimensionality on one hand and by 
the weak $\bm{q}$-dependence on the other hand.

\begin{figure}[t]
\begin{center}
\includegraphics[width=8.0cm]{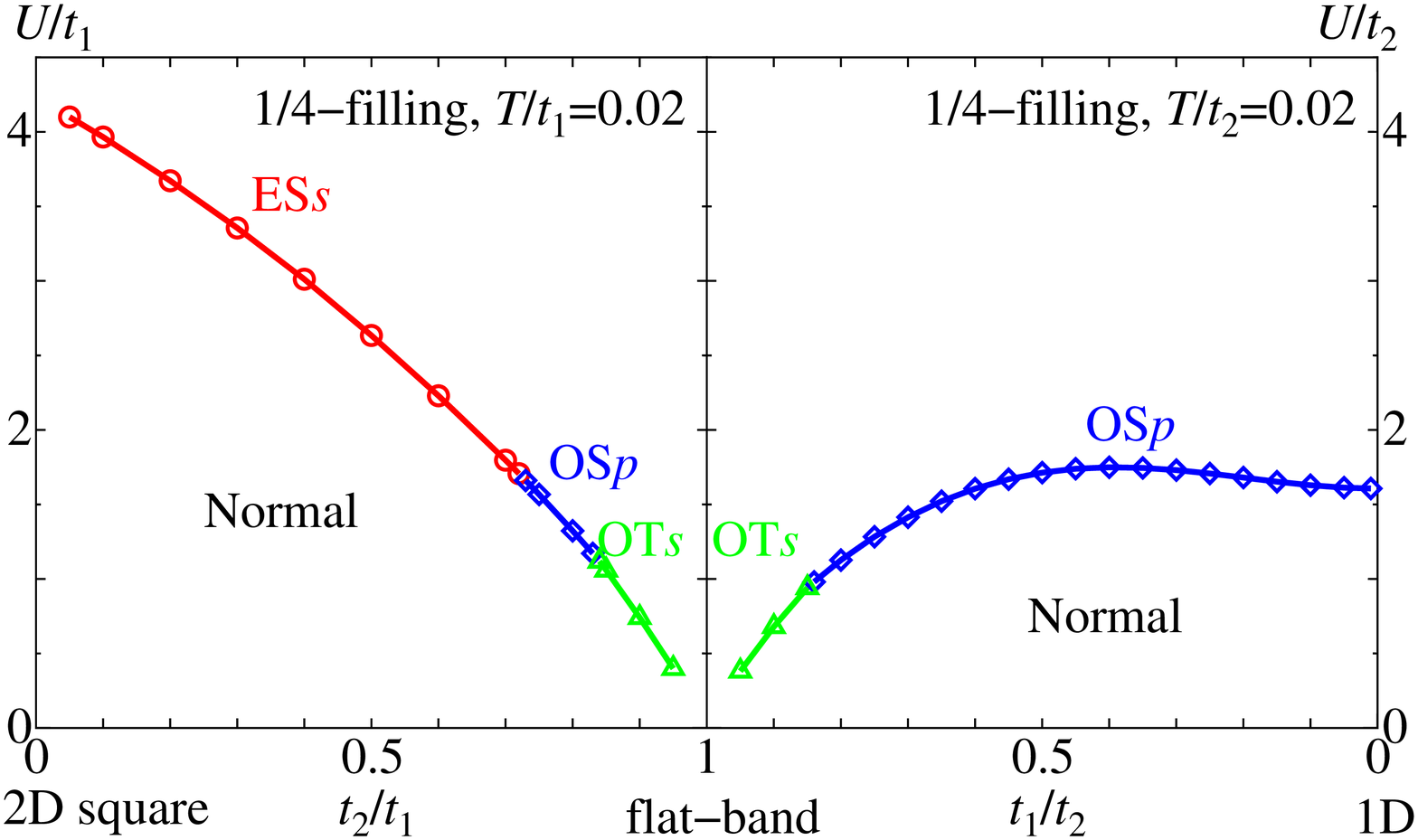}
\caption{(Color online) Phase diagram on the $t_1$- and $t_2$-$U$
 planes, where the open circles, diamonds and triangles denote the ES$s$,
 OS$p$ and OT$s$ pairing instabilities, respectively. \label{fig_pd}}
\end{center}
\end{figure}

Finally, we show the superconducting phase diagram 
in the $t_1$- and $t_2$-$U$ planes in
Fig. \ref{fig_pd},
where the superconducting phase boundary are defined as
the points at which the largest eigenvalue of eq. (\ref{eq_gap}) $\lambda$ reaches unity. 
For $t_1>t_2$, the ES$s$ pairing is realized in the wide range of $t_2/t_1$ near the
magnetic ordered phase with the ordering vector 
$\bm{q}\sim (\pi/2,\pi/2)$ [see Figs. \ref{fig_chi0} (d) and (e) and
Fig. \ref{fig_pd0}]. 
 On the other hand, for $t_1<t_2$, the OS$p$ state is
realized in the extremely wide range of $t_1/t_2$.  
   This situation is in a striking contrast to the usual
quasi-one-dimensional Hubbard model, where the
one-dimensionality is lost rapidly with increasing inter-chain hoppings
leading to  suppression of  the OS$p$ pairing correlation. 
The OS$p$ pairing is realized also for $t_1>t_2$ ($0.73 \le t_2/t_1 \le
0.83$) 
because there the 1D structure of
$\hat{\chi}^{(0)}(q)$ develops as shown in Fig. \ref{fig_chi0}
(e). 
  Remarkably, the OS$p$ pairing emerges already 
  for moderate enhancement of the spin fluctuations, 
 whereas the ES$s$ pairing necessitates the
  very strong enhancement, which indicates the robustness of
  the OS$p$ pairing\cite{shigeta}. 
 It is worthwhile to note that for $t_1<t_2$, 
the ES$s$ state is second dominant in the
wide range of $t_1/t_2$. 
The gap function for this ES$s$ pairing, 
however, have no nodes on the Fermi surfaces 
in contrast to the
case that $t_1>t_2$ because the Fermi surfaces are
disconnected\cite{kuroki}. It is similar to the  
$s_{\pm}$-wave state discussed for the iron-based
superconductors\cite{mazin,kuroki_2,yanagi}. 
At $1/4$-filling, the Fermi surface around
$\bm{k}=(\pi,\pi)$ is smaller than that around $\bm{k}=(0,0)$ and 
the resulting $T_c$ for the ES$s$ state is relatively low, 
while near $1/2$-filling, the Fermi surfaces around $\bm{k}=(\pi,\pi)$ 
is comparable in size as that around
$\bm{k}=(0,0)$, and the resulting $T_c$ is relatively high (not shown)\cite{tbp}. 
Near the Mielke point, the OT$s$ state is realized due to the spin
fluctuations with featureless $\bm{q}$-dependence. 
 It should be pointed out that when the $\bm{q}$-dependence of the spin
fluctuations are weak, the mode-mode coupling effects of the various
fluctuations are considered to become important. 
Thus, the vertex corrections as well as 
the self-energy corrections\cite{isoda} neglected in the RPA may play 
significant roles for the stability of the OT$s$ pairing near the
Mielke point.
We note that with
decreasing temperature, the OT$s$ region shrinks and the ES$s$ region
for $t_1>t_2$ and
the OS$p$ region for $t_1<t_2$ are extended toward the Mielke
point because the $\bm{q}$-dependence of the spin fluctuations gets
stronger. Therefore, further investigations beyond the RPA study are needed to
clarify whether the OT$s$ pairing near the Mielke point eventually survives at lower
temperatures.

Here, we briefly comment on the re-entrant feature of the odd-frequency
pairing previously reported\cite{fuseya,kusunose}. 
In the present study, we have not observed any clear re-entrant feature
down to $T=0.01t_1$ or $T=0.01t_2$.

In summary, we have investigated the magnetic properties and the
superconductivity in the CB-lattice Hubbard model at $1/4$-filling
with use of the mean field approximation, ED method and RPA.  
We have shown that the model exhibits the
one-dimensional charge and magnetic orders, 
such as the plaquette and line orders. 
The spin fluctuations have also the one-dimensional feature, i.e., the
ridge-like structures in the momentum space and 
drives the OS$p$ superconductivity in the extremely wide range of
$t_1/t_2$.  
These phenomena are due to the 1D nature of the
CB-lattice Hubbard model which is 
quite robust against the inter-chain hopping $t_1$ 
in contrast to the case of the usual quasi-1D Hubbard model.

\begin{acknowledgment}
This work has been supported by a Grant-in-Aid for
Scientific Research on Innovative Areas ``Heavy
Electrons'' (No. 20102008) of The Ministry of
Education, Culture, Sports, Science, and Technology,
Japan.
\end{acknowledgment}

\end{document}